\documentclass[preprint,12pt]{elsarticle}

\usepackage{color}
\usepackage{graphicx}
\usepackage{amssymb}
\usepackage{amsfonts}
\usepackage{amsmath}
\usepackage{mdframed}

\usepackage{xspace}
\usepackage{fancyvrb}
\usepackage{placeins}
\usepackage{longtable,multirow}
\usepackage{hyperref}
\usepackage{url}

\usepackage{blindtext}
\usepackage{enumitem}

\usepackage[toc,page]{appendix}
\usepackage{lineno}

\newcounter{bla}

\newcommand{\smo}{\textsf{SModelS}}
\newcommand{\json}{\textsf{JSON}}
\newcommand{\pyhf}{\textsf{pyhf}}
\newcommand{\python}{\textsf{python}}

\newcommand{\Ztwo}{$\mathbb{Z}_2$}
\journal{arXiv}
\begin{document}

%=========================================================================
\begin{frontmatter}
%=========================================================================

\title{A SModelS interface for pyhf likelihoods}

\author[a]{Ga\"{e}l Alguero}
\author[a]{Sabine Kraml}
\author[b,c]{Wolfgang Waltenberger}
\cortext[x]{\textit{E-mail addresses:} gael.alguero@lpsc.in2p3.fr, sabine.kraml@lpsc.in2p3.fr,\\ wolfgang.waltenberger@oeaw.ac.at}

\address[a]{Laboratoire de Physique Subatomique et de Cosmologie, Universit\'e
  Grenoble-Alpes, CNRS/IN2P3, 53 Avenue des Martyrs, F-38026 Grenoble, France}
\address[b]{Institut f\"ur Hochenergiephysik,  \"Osterreichische Akademie der
  Wissenschaften, Nikolsdorfer Gasse 18, 1050 Wien, Austria}
\address[c]{University of Vienna, Faculty of Physics, Boltzmanngasse 5, A-1090 Wien, Austria}

\begin{abstract} 
\smo\ is an automatized tool enabling the fast interpretation of simplified model results from the LHC within any model of new physics respecting a \Ztwo\ symmetry.  We here present a new version of \smo, which can use the full likelihoods now provided by ATLAS in the form of \pyhf\ \json\ files. This much improves the statistical evaluation and therefore also the limit setting on new physics scenarios.
\end{abstract}


\end{frontmatter}

%\linenumbers

%% CPC publications require a program summary
%\input{prog_summary}

%=========================================================================
\section{Introduction}\label{intro}
%=========================================================================

An essential step for interpretation of experimental results is the construction of a statistical model, or \emph{likelihood}, to compare the observed data to the target theory. Given the likelihood, all the standard statistical approaches are available for extracting information from it. 

Therefore, Ref.~\cite{Kraml:2012sg} recommended for the presentation of LHC results: 
{\it ''When feasible, provide a mathematical description of the final likelihood function in which experimental data and parameters are clearly  distinguished,  either  in  the  publication  or  the  auxiliary  information. Limits of validity should always be clearly specified.''}
And furthermore {\it ``Additionally  provide  a  digitized  implementation  of the likelihood that is consistent with the mathematical description.''}
These are the Les Houches Recommendations 3(b) and 3(c). 
The necessity of detailed likelihood information was further  elaborated in the recent report of the LHC Reinterpretation Forum~\cite{Abdallah:2020pec}. 

Among the major benefits of detailed likelihood information for reinterpretation is the fact that it allows one to statistically  combine disjoint signal regions (SRs) instead of using only the most sensitive (a.k.a.\ ``best'') SR; 
see, e.g., \cite{Asadi:2017qon,Athron:2018vxy} for the impact in  physics studies.

The CMS SUSY group has been publishing SR correlation data in the form of covariance matrices for some of their analyses.
This so-called simplified likelihood~\cite{CMS:2242860} approach assumes that uncertainties can be well approximated by Gaussians.   
\smo~\cite{Kraml:2013mwa,Ambrogi:2017neo,Ambrogi:2018ujg} can make use of these correlation data since its version 1.2~\cite{Ambrogi:2018ujg}; their benefit for limit setting 
was  demonstrated in \cite{Ambrogi:2018ujg} and contribution~15 of \cite{Brooijmans:2020yij}.%
\footnote{Non-Gaussian effects can also be incorporated in the simplified likelihood framework. To this end,  Ref.~\cite{Buckley:2018vdr} proposed a simple method to encode asymmetry information into correlations via publication of only $N_\text{bins}$ additional numbers (as opposed to the more common $N_\text{bins} \times N_\text{bins}$ second order correlation data).}

ATLAS has recently gone a significant step further by publishing 
\emph{full likelihoods} using a \json\ serialization~\cite{ATL-PHYS-PUB-2019-029}, which provides background estimates, changes under systematic variations, and observed data counts at the same fidelity as used in the experiment.
The \json\ format describes the \textsf{HistFactory} family of statistical models~\cite{Cranmer:1456844}, which is used by the majority of ATLAS searches.  The \pyhf\ package~\cite{pyhf} is then used to construct statistical models, and perform statistical inference, within a \python\ environment. 
Note that this fulfills for the first time the Les Houches Recommendations~3(b,c)!  

In the following we describe the usage of the ATLAS \pyhf\   likelihoods in \smo. 
We also demonstrate the improvements in the statistical evaluation ---and thus in the constraining power---due to these likelihoods. 
Readers who are not already familiar with \smo\ are referred to
\cite{Kraml:2013mwa,Ambrogi:2017neo,Ambrogi:2018ujg,Dutta:2018ioj,Khosa:2020zar} for details on the tool and how to use it. Further information, including a detailed online manual, is available at \url{https://smodels.github.io/}.

%=========================================================================
\section{Usage in SModelS}\label{interface}
%=========================================================================

The \pyhf\ \json\ files~\cite{ATL-PHYS-PUB-2019-029} from ATLAS 
report $L(\theta|\mathcal{D})$, where $\theta$ is the union of parameters of interest and possible nuisance parameters and $\mathcal{D}$ denotes the observed data. 
Encoded in this way are, in particular, background estimates, correlations, and primary data. %, as well as simplified model estimates. 
Together with the relevant simplified model efficiency maps, they allow \smo\ to evaluate the likelihood of the signal strength of a hypothesized signal in a realistic manner.  

To make use of this %information
machinery, besides {\tt scipy} and {\tt numpy}, which are already required by \smo, the following 
python packages need to be installed: 
%\begin{quote}
%  {\tt pyhf>=0.5.1} ({\tt >=0.5.2} recommended),\\
%  %{\tt json>=2.0.9}, 
%  {\tt jsonpatch}, 
%  {\tt jsonschema>=3.2.0},
%  {\tt torch>=1.6.0}.
%\end{quote}
\begin{quote}
 {\tt pyhf}, {\tt jsonpatch}, {\tt jsonschema}.  
\end{quote}
In addition, for speed reasons, we recommend {\tt pytorch} as backend for {\tt pyhf} (if not available, the default backend will be used). 
Details are given in the online manual at 
\url{https://smodels.readthedocs.io/en/stable/Installation.html}. 
Details on the \pyhf\ package are given in \cite{pyhf} 
and at \url{https://scikit-hep.org/pyhf/}.

\subsection{Implementation in the database}

In the \smo\ database, the \json\ files are placed in the respective analysis folder that holds the simplified model efficiency maps (see \cite{Ambrogi:2017neo} for the database structure). The information, which \json\ file is used to combine which SRs, is given in the {\tt globalInfo.txt} file in each analysis folder. For example, for the ATLAS stau search~\cite{Aad:2019byo}, 
which has two SRs, the {\tt globalInfo.txt} file contains: 

\begin{verbatim}
  id: ATLAS-SUSY-2018-04
  ....
  datasetOrder: "SRlow", "SRhigh"
  jsonFiles: {"SRcombined.json": ["SRlow", "SRhigh"]}    
\end{verbatim}

\noindent 
In case the provided \json\ files describe the combination of one or more subsets of SRs, as in the multi-$b$ sbottom search~\cite{Aad:2019pfy}, the format is: 

\begin{verbatim}
  id: ATLAS-SUSY-2018-31
  ....
  datasetOrder: "SRA_L", "SRA_M", "SRA_H", "SRB", "SRC_22", 
                "SRC_24", "SRC_26", "SRC_28"
  jsonFiles: {"BkgOnlyA.json": ["SRA_L", "SRA_M", "SRA_H"],
              "BkgOnlyB.json": ["SRB"], 
              "BkgOnlyC.json": ["SRC_22", "SRC_24", "SRC_26",
                                "SRC_28"]}
\end{verbatim}
Here, the likelihoods for SRs A, B and C will first be  evaluated separately, and then only the most sensitive result among SRA, SRB and SRC will be used for the limit setting.

\subsection{Changes/additions in the SModelS code}

The interfacing of \pyhf\ to \smo\ can be summarized in two parts: the addition of an independent module {\tt tools/pyhfInterface.py}, and the changes brought to {\tt experiment/datasetObj.py}.

The {\tt tools/pyhfInterface.py} module is made of two classes, {\tt PyhfData}, storing and handling informations related to the \json\ files and input signal predictions, and {\tt PyhfUpperLimitComputer}, where the upper limits are inferred given the {\tt PyhfData} information.
The constructor of {\tt PyhfData} takes as arguments {\tt nsignals} and {\tt inputJsons}, which are respectively the list of BSM prediction yields and the list of workspaces, i.e., the likelihoods as \python\ \json\ objects \cite{pezoa2016foundations}. The list of signal yields is a 2-dimensional list, so that there is a sublist for each \json\ likelihood. For the previous example
\begin{verbatim}
  jsonFiles: {"BkgOnlyA.json": ["SRA_L", "SRA_M", "SRA_H"],
              "BkgOnlyB.json": ["SRB"], 
              "BkgOnlyC.json": ["SRC_22", "SRC_24", "SRC_26",
                                "SRC_28"]}
\end{verbatim}
the {\tt nsignals} would read
\begin{verbatim}
  nsignals = [[<SRA_L>, <SRA_M>, <SRA_H>],
              [<SRB>], 
              [<SRC_22>, <SRC_24>, <SRC_26>, <SRC_28>]]
\end{verbatim}
where {\verb <SRA_L> }, {\verb <SRA_M> }, ...\ are the event yield predictions in the signal regions named {\verb "SRA_L" }, {\verb "SRA_M" }, ..., respectively.

The \json\ likelihoods provided by ATLAS are written in the following \python\ dictionary structure: 
\begin{verbatim}
    {"channels":[
        {"name":..., "samples":[
            {"data":[...], "modifiers":[...]},
            {"data":[...], "modifiers":[...]},
            ...
            ]
        },
        {"name":..., "samples":[...],
        ...
        ]
    }
\end{verbatim}
where the {\tt channels} are the usual signals regions, and the {\tt samples} contain the different background  contributions. %; as well as the signal yields is written as one of these {\tt samples}. 
In each sample, {\tt data} contains the event yields and {\tt modifiers} is the list of all the modifiers representing the uncertainties. The hypothesized BSM signal will be added in the form of one of these {\tt samples}.

The {\tt PyhfData} constructor first collects information in the workspaces such as the number of SRs, and the paths to the {\tt samples} where the BSM predictions are to be written, and also the virtual regions (VRs) and control regions (CRs) that are assumed not to contribute and are then removed from the workspaces. It must be noted that this approximation can imply a slight loss in accuracy because any potential leakage of the signal into the VRs or CRs is neglected. The fetched information in the {\tt inputJsons} is then compared to the {\tt nsignals} to check for any inconsistencies in the format of the two variables.

The {\tt jsonpatch} package \cite{Stefan2020jsonpatch} allows to easily write into an existing \json\ object. The {\tt PyhfUpperLimitComputer} class uses this feature to add the BSM prediction yields and remove the control and virtual regions from the workspaces. This procedure is dynamical so that the signal predictions can be re-scaled throughout the statistical inference.

The {\tt pyhf.infer.hypotest} allows to compute the $\textrm{CL}_s$~\cite{Read:2002hq} with a signal strength modifier $\mu$ as argument, using the asymptotic formulae from~\cite{Cowan_2011}.
Upper limits are found by varying the $\textrm{CL}_s$ with respect to $\mu$. Namely, our \pyhf\ interface will look for the $\mu$ at $95\%$ exclusion confidence level (CL). $\mu$ being a multiplicative factor, the unit of the obtained upper limit will depend on the unit of the signal predictions provided. In our case, normalised signals give unitless upper limits on the event yields. We first dynamically rescale the signal predictions, so that $\mu$ at $95\%$ CL lies in the interval $[0.2,\, 5]$, and then use the {\tt optimize} feature of the {\tt scipy} package \cite{2020SciPy-NMeth} to find the exclusion limit at $95\%$~CL. 

The independent {\tt tools/pyhfInterface.py} module is interfaced to \smo\ in {\tt experiment/datasetObj.py}, as it is for the simplified likelihood. If combination is requested and \json\ files are found in the database, the code in {\tt datasetObj.py} will perform \pyhf\ combination. If more than one \json\ file is provided, "best expected combination" is performed, i.e., the upper limit is computed using the \json\ that gives the most sensitive combination.

\subsection{Running SModelS}

The interface to \pyhf\ is available from \smo \,v1.2.4 onward. Running the program has not changed with respect to previous versions, apart from setting a switch to evoke the (optional) use of the \json\ files in the database. 
When using {\tt runSModelS.py}, one has to set 
\begin{verbatim}
   combineSRs = True
\end{verbatim}
in the {\tt paramerers.ini} file. % to evoke the use of the \json\ files in the database. 
Note that the same flag also turns on the SR combination 
in the simplified likelihood approach for CMS efficiency map results, for which a covariance matrix is available. 

Alternatively, one can call {\tt theoryPredictionsFor()} with the option {\tt combinedResults=True} in one's own \python\ program, cf.\ the {\tt Example.py} file in the \smo \,v1.2.4 distribution.

%=========================================================================
\section{Validation and physics impact}\label{validation}
%=========================================================================

\begin{figure}[t]\centering
\includegraphics[width=0.5\textwidth]{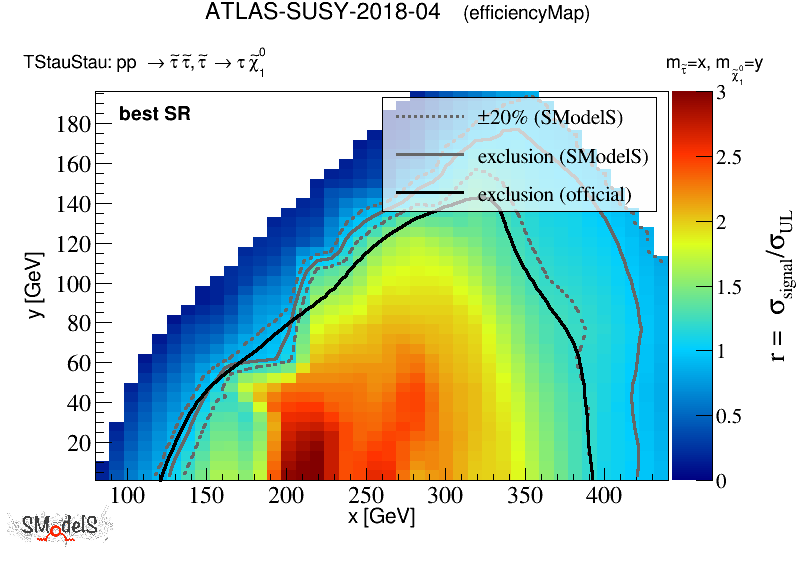}%
\includegraphics[width=0.5\textwidth]{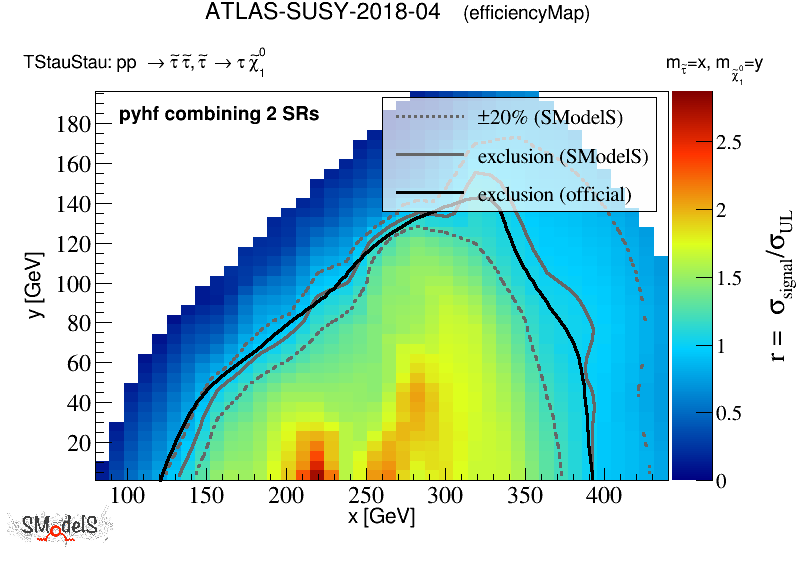}
\caption{Validation of TStauStau ($pp\to \tilde\tau_1^+\tilde\tau_1^-$, $\tilde\tau^\pm_1\to\tau \tilde\chi^0_1$) result from the ATLAS stau search~\cite{Aad:2019byo}, on the left using the best SR, on the right using the full likelihood.}
\label{fig:ATLAS-SUSY-2018-04}
\end{figure}

We compare in Figure~\ref{fig:ATLAS-SUSY-2018-04} the \smo\ exclusion (grey line) with the official exclusion (black line) for the ATLAS stau search~\cite{Aad:2019byo}, using best SR (left) and using \pyhf\ combination (right). As one can see, the usual procedure, which picks up the most sensitive efficiency map result, over-excludes by about 50 GeV on half the exclusion line. In contrast, a very good agreement with the official ATLAS result is obtained with the full \pyhf\ likelihood.\footnote{The remaining small difference might be due to the (interpolated) acceptance $\times$ efficiency values from the simplified model efficiency maps not exactly matching the ``true'' ones of the experimental analysis.}  

\begin{figure}[t]\centering
\includegraphics[width=0.5\textwidth]{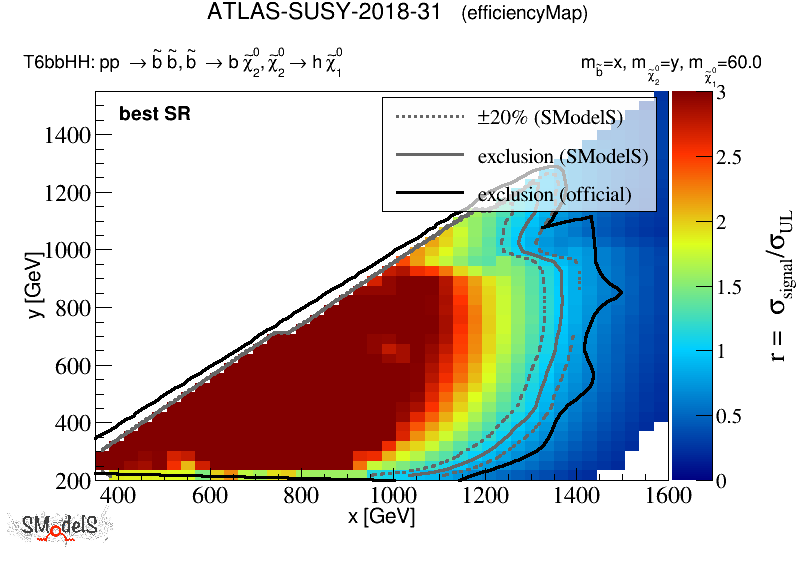}%
\includegraphics[width=0.5\textwidth]{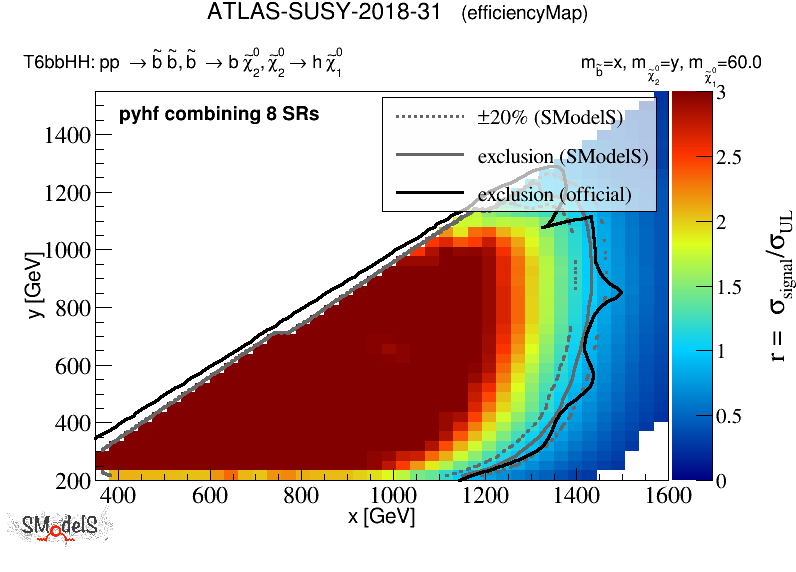}
\caption{Validation of the T6bbHH ($pp\to\tilde b_1^{}\tilde b_1^*$, $\tilde b_1^{}\to b\tilde\chi^0_2$, $\tilde\chi^0_2\to h\tilde\chi^0_1$) result from the ATLAS sbottom search~\cite{Aad:2019pfy}, on the left using the best SR, on the right using the full likelihood.}
\label{fig:ATLAS-SUSY-2018-31dm60}
\end{figure}

Figure~\ref{fig:ATLAS-SUSY-2018-31dm60} shows the same kind of validation for the ATLAS sbottom search~\cite{Aad:2019pfy}, which was actually the first one to provide the full likelihood. In this case, without \pyhf, \smo\ is under-excluding by roughly 50--100 GeV.\footnote{This under-exclusion is even more pronounced when using the inclusive instead of the exclusive SRs for this analysis.}  Again we observe a significant improvement with the \pyhf\ combination. 

\begin{figure}[t]\centering
\includegraphics[width=0.5\textwidth]{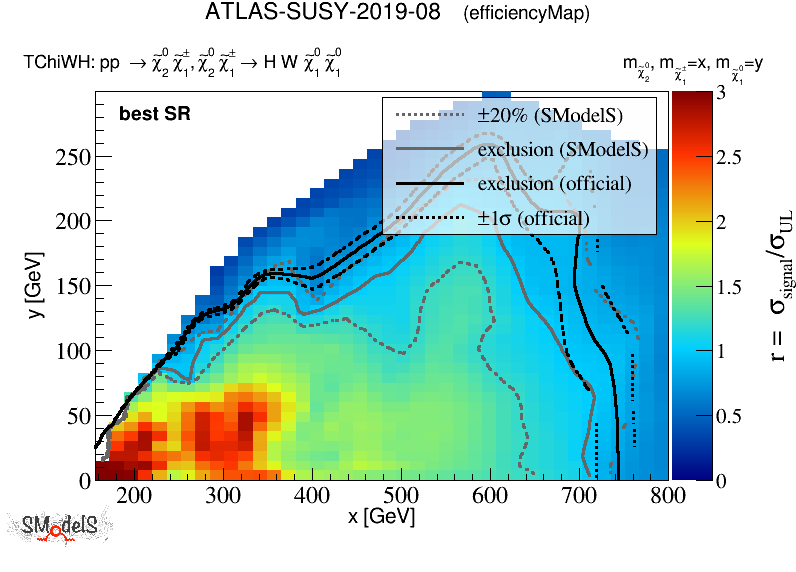}%
\includegraphics[width=0.5\textwidth]{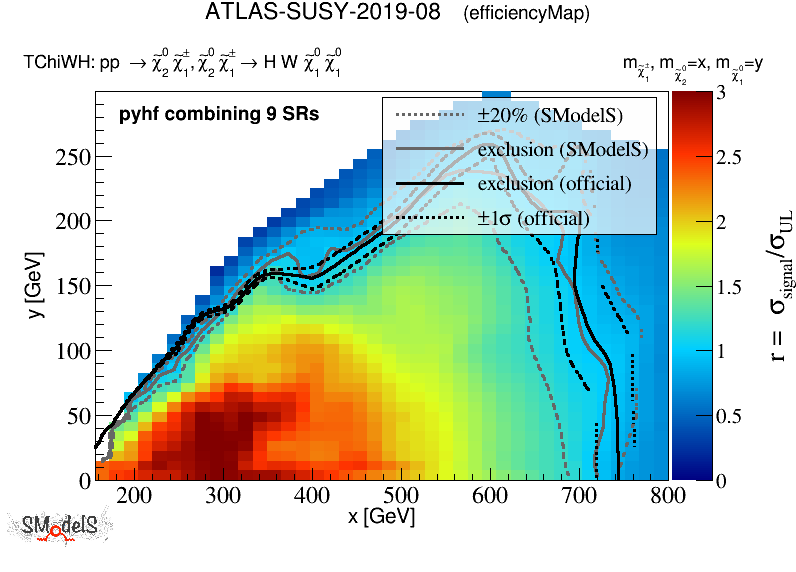}
\caption{Validation of the TChiWH ($pp\to\tilde \chi_2^{0}\tilde \chi_1^{\pm}$, $\tilde \chi_2^{0}\to h\tilde\chi^0_1$, $\tilde\chi^{\pm}_1\to W\tilde\chi^0_1$) result from the ATLAS electroweakino search~\cite{Aad:2019vvf}, on the left using the best SR, on the right using the full likelihood.}
\label{fig:ATLAS-SUSY-2019-08}
\end{figure}

Our third example, shown in Figure~\ref{fig:ATLAS-SUSY-2019-08}, is for the ATLAS electroweakino search in the $W(\to \ell\nu) h(\to b\bar b)+E_T^{\rm miss}$ channel \cite{Aad:2019vvf}. 
Using the best exclusive SR (left panel in Figure~\ref{fig:ATLAS-SUSY-2019-08}), we face an under-exclusion over almost the entire mass plane. Using instead the best inclusive SR (not shown) would give a \smo\ limit closer to the official one for large mass differences, but lead to a serious over-exclusion for small mass differences. The combination of SRs based on the full likelihood resolves these problems, and we obtain a good agreement of the \smo\ exclusion line with the official one from ATLAS as shown in the right panel of 
Figure~\ref{fig:ATLAS-SUSY-2019-08}. 

Even though we only show three results here, one can appreciate the gain in accuracy one can reach with using \pyhf\ and full likelihoods. The ATLAS collaboration is at the beginning of a huge effort to provide  full statistical models for new analyses. The first analyses published already show how this can help theorists make more trustful reinterpretations. The importance of such likelihood information for, e.g., global fits, has also been emphasised in~\cite{Abdallah:2020pec}. 

%\clearpage
%=========================================================================
\section{Conclusions}\label{conclusions}
%=========================================================================

We presented an interface of \smo\ to \pyhf\ that enables the use 
of the full likelihoods provided by ATLAS in the form of \pyhf\ \json\ files. 
The \smo\ database was extended by efficiency map results with the corresponding \json\ files of three new ATLAS SUSY analyses \cite{Aad:2019byo,Aad:2019pfy,Aad:2019vvf} 
for full Run~2 luminosity (139~fb$^{-1}$). 

The new version, \smo\,v1.2.4, 
is publicly available from \url{https://smodels.github.io/}  
and can readily be employed for physics studies.
We congratulate ATLAS to the important move of making full likelihood information available in digital format and are looking forward to including more such data in future updates of \smo.

This completes the work started in contribution~15 of \cite{Brooijmans:2020yij} for \smo; the \textsf{MadAnalysis\,5} interface to \pyhf\ should become available in the upcoming \textsf{MadAnalysis\,5} v1.9 release. 

Last but not least we note that the technical discussions with the \pyhf\ team are handled via github's issue tracking system, see e.g.\  \url{https://github.com/scikit-hep/pyhf/issues/620}, and are thus transparent and open to all.
 
%=========================================================================
\section*{Acknowledgements}
%=========================================================================

We thank ATLAS for the important step to make full likelihood information available on \textsf{HEPData}. Our special thanks also go to Matthew Feickert, Lukas Heinrich, and Giordon Stark for technical help with \pyhf.  Finally, we are indebted to Andre Lessa for dedicated checks and help with the public release of \smo\,v1.2.4.
\paragraph{Funding:} The work of G.A.\ and S.K.\ was supported in part by the IN2P3 project ``Th\'eorie -- BSMGA''.

%=========================================================================
\section*{References}
%=========================================================================
\bibliographystyle{elsarticle-num}
\bibliography{references}

\begin{thebibliography}{10}
\expandafter\ifx\csname url\endcsname\relax
  \def\url#1{\texttt{#1}}\fi
\expandafter\ifx\csname urlprefix\endcsname\relax\def\urlprefix{URL }\fi
\expandafter\ifx\csname href\endcsname\relax
  \def\href#1#2{#2} \def\path#1{#1}\fi

\bibitem{Kraml:2012sg}
S.~Kraml, et~al., {Searches for New Physics: Les Houches Recommendations for
  the Presentation of LHC Results}, Eur. Phys. J. C 72 (2012) 1976.
\newblock \href {http://arxiv.org/abs/1203.2489} {\path{arXiv:1203.2489}},
  \href {http://dx.doi.org/10.1140/epjc/s10052-012-1976-3}
  {\path{doi:10.1140/epjc/s10052-012-1976-3}}.

\bibitem{Abdallah:2020pec}
W.~Abdallah, et~al., {Reinterpretation of LHC Results for New Physics: Status
  and Recommendations after Run 2}, SciPost Phys. 9 (2020) 22.
\newblock \href {http://arxiv.org/abs/2003.07868} {\path{arXiv:2003.07868}},
  \href {http://dx.doi.org/10.21468/SciPostPhys.9.2.022}
  {\path{doi:10.21468/SciPostPhys.9.2.022}}.

\bibitem{Asadi:2017qon}
P.~Asadi, M.~R. Buckley, A.~DiFranzo, A.~Monteux, D.~Shih, {Digging Deeper for
  New Physics in the LHC Data}, JHEP 11 (2017) 194.
\newblock \href {http://arxiv.org/abs/1707.05783} {\path{arXiv:1707.05783}},
  \href {http://dx.doi.org/10.1007/JHEP11(2017)194}
  {\path{doi:10.1007/JHEP11(2017)194}}.

\bibitem{Athron:2018vxy}
P.~Athron, et~al., {Combined collider constraints on neutralinos and
  charginos}, Eur. Phys. J. C 79~(5) (2019) 395.
\newblock \href {http://arxiv.org/abs/1809.02097} {\path{arXiv:1809.02097}},
  \href {http://dx.doi.org/10.1140/epjc/s10052-019-6837-x}
  {\path{doi:10.1140/epjc/s10052-019-6837-x}}.

\bibitem{CMS:2242860}
{CMS Collaboration}, {Simplified likelihood for the re-interpretation of public
  CMS results}, Tech. Rep. CMS-NOTE-2017-001, CERN, Geneva,
  \url{https://cds.cern.ch/record/2242860} (Jan 2017).

\bibitem{Kraml:2013mwa}
S.~Kraml, S.~Kulkarni, U.~Laa, A.~Lessa, W.~Magerl, D.~Proschofsky,
  W.~Waltenberger, {SModelS: a tool for interpreting simplified-model results
  from the LHC and its application to supersymmetry}, Eur.Phys.J. C74 (2014)
  2868.
\newblock \href {http://arxiv.org/abs/1312.4175} {\path{arXiv:1312.4175}},
  \href {http://dx.doi.org/10.1140/epjc/s10052-014-2868-5}
  {\path{doi:10.1140/epjc/s10052-014-2868-5}}.

\bibitem{Ambrogi:2017neo}
F.~Ambrogi, S.~Kraml, S.~Kulkarni, U.~Laa, A.~Lessa, V.~Magerl, J.~Sonneveld,
  M.~Traub, W.~Waltenberger, {SModelS v1.1 user manual: Improving simplified
  model constraints with efficiency maps}, Comput. Phys. Commun. 227 (2018)
  72--98.
\newblock \href {http://arxiv.org/abs/1701.06586} {\path{arXiv:1701.06586}},
  \href {http://dx.doi.org/10.1016/j.cpc.2018.02.007}
  {\path{doi:10.1016/j.cpc.2018.02.007}}.

\bibitem{Ambrogi:2018ujg}
F.~Ambrogi, et~al., {SModelS v1.2: long-lived particles, combination of signal
  regions, and other novelties}, Comput. Phys. Commun. 251 (2020) 106848.
\newblock \href {http://arxiv.org/abs/1811.10624} {\path{arXiv:1811.10624}},
  \href {http://dx.doi.org/10.1016/j.cpc.2019.07.013}
  {\path{doi:10.1016/j.cpc.2019.07.013}}.

\bibitem{Brooijmans:2020yij}
G.~Brooijmans, et~al., {Les Houches 2019 Physics at TeV Colliders: New Physics
  Working Group Report}, in: {11th Les Houches Workshop on Physics at TeV
  Colliders}: {PhysTeV Les Houches}, 2020.
\newblock \href {http://arxiv.org/abs/2002.12220} {\path{arXiv:2002.12220}}.

\bibitem{Buckley:2018vdr}
A.~Buckley, M.~Citron, S.~Fichet, S.~Kraml, W.~Waltenberger, N.~Wardle, {The
  Simplified Likelihood Framework}, JHEP 04 (2019) 064.
\newblock \href {http://arxiv.org/abs/1809.05548} {\path{arXiv:1809.05548}},
  \href {http://dx.doi.org/10.1007/JHEP04(2019)064}
  {\path{doi:10.1007/JHEP04(2019)064}}.

\bibitem{ATL-PHYS-PUB-2019-029}
{ATLAS Collaboration}, {Reproducing searches for new physics with the ATLAS
  experiment through publication of full statistical likelihoods}, Tech. Rep.
  ATL-PHYS-PUB-2019-029, CERN, Geneva, \url{https://cds.cern.ch/record/2684863}
  (Aug 2019).

\bibitem{Cranmer:1456844}
K.~Cranmer, G.~Lewis, L.~Moneta, A.~Shibata, W.~Verkerke, {HistFactory: A tool
  for creating statistical models for use with RooFit and
  RooStats}~(CERN-OPEN-2012-016), \url{https://cds.cern.ch/record/1456844}.

\bibitem{pyhf}
L.~Heinrich, M.~Feickert, G.~Stark, {scikit-hep/pyhf},
  \url{https://github.com/scikit-hep/pyhf}, DOI:
  \href{https://doi.org/10.5281/zenodo.1169739}{10.5281/zenodo.1169739}.
\newblock \href {http://dx.doi.org/10.5281/zenodo.1169739}
  {\path{doi:10.5281/zenodo.1169739}}.

\bibitem{Dutta:2018ioj}
J.~Dutta, S.~Kraml, A.~Lessa, W.~Waltenberger, {SModelS extension with the CMS
  supersymmetry search results from Run 2}, LHEP 1~(1) (2018) 5--12.
\newblock \href {http://arxiv.org/abs/1803.02204} {\path{arXiv:1803.02204}},
  \href {http://dx.doi.org/10.31526/LHEP.1.2018.02}
  {\path{doi:10.31526/LHEP.1.2018.02}}.

\bibitem{Khosa:2020zar}
C.~K. Khosa, S.~Kraml, A.~Lessa, P.~Neuhuber, W.~Waltenberger, {SModelS
  database update v1.2.3}, {to appear in LHEP. }\href
  {http://arxiv.org/abs/2005.00555} {\path{arXiv:2005.00555}}.

\bibitem{Aad:2019byo}
G.~Aad, et~al., {Search for direct stau production in events with two hadronic
  $\tau$-leptons in $\sqrt{s} = 13$ TeV $pp$ collisions with the ATLAS
  detector}, Phys. Rev. D 101~(3) (2020) 032009.
\newblock \href {http://arxiv.org/abs/1911.06660} {\path{arXiv:1911.06660}},
  \href {http://dx.doi.org/10.1103/PhysRevD.101.032009}
  {\path{doi:10.1103/PhysRevD.101.032009}}.

\bibitem{Aad:2019pfy}
G.~Aad, et~al., {Search for bottom-squark pair production with the ATLAS
  detector in final states containing Higgs bosons, $b$-jets and missing
  transverse momentum}, JHEP 12 (2019) 060, {HEPData entry}:
  \url{https://doi.org/10.17182/hepdata.89408}.
\newblock \href {http://arxiv.org/abs/1908.03122} {\path{arXiv:1908.03122}},
  \href {http://dx.doi.org/10.1007/JHEP12(2019)060}
  {\path{doi:10.1007/JHEP12(2019)060}}.

\bibitem{pezoa2016foundations}
F.~Pezoa, J.~L. Reutter, F.~Suarez, M.~Ugarte, D.~Vrgo{\v{c}}, Foundations of
  json schema, in: Proceedings of the 25th International Conference on World
  Wide Web, International World Wide Web Conferences Steering Committee, 2016,
  pp. 263--273.

\bibitem{Stefan2020jsonpatch}
S.~K\"{o}gl, \href{https://github.com/stefankoegl/python-json-patch}{jsonpatch}
  (6 2020).
\newline\urlprefix\url{https://github.com/stefankoegl/python-json-patch}

\bibitem{Read:2002hq}
A.~L. Read, {Presentation of search results: The CL(s) technique}, J. Phys. G
  28 (2002) 2693--2704.
\newblock \href {http://dx.doi.org/10.1088/0954-3899/28/10/313}
  {\path{doi:10.1088/0954-3899/28/10/313}}.

\bibitem{Cowan_2011}
G.~Cowan, K.~Cranmer, E.~Gross, O.~Vitells,
  \href{http://dx.doi.org/10.1140/epjc/s10052-011-1554-0}{Asymptotic formulae
  for likelihood-based tests of new physics}, The European Physical Journal C
  71~(2).
\newblock \href {http://dx.doi.org/10.1140/epjc/s10052-011-1554-0}
  {\path{doi:10.1140/epjc/s10052-011-1554-0}}.
\newline\urlprefix\url{http://dx.doi.org/10.1140/epjc/s10052-011-1554-0}

\bibitem{2020SciPy-NMeth}
P.~{Virtanen}, R.~{Gommers}, T.~E. {Oliphant}, et~al., {SciPy 1.0: Fundamental
  Algorithms for Scientific Computing in Python}, Nature Methods 17 (2020)
  261--272.
\newblock \href {http://dx.doi.org/https://doi.org/10.1038/s41592-019-0686-2}
  {\path{doi:https://doi.org/10.1038/s41592-019-0686-2}}.

\bibitem{Aad:2019vvf}
G.~Aad, et~al., {Search for direct production of electroweakinos in final
  states with one lepton, missing transverse momentum and a Higgs boson
  decaying into two $b$-jets in (pp) collisions at $\sqrt{s}=13$ TeV with the
  ATLAS detector}, Eur. Phys. J. C 80~(8) (2020) 691, {HEPData entry}:
  \url{https://doi.org/10.17182/hepdata.90607.v2}.
\newblock \href {http://arxiv.org/abs/1909.09226} {\path{arXiv:1909.09226}}.

\end{thebibliography}

\end{document}